# Generation of photoluminescent ultrashort carbon nanotubes through nanoscale exciton localization at sp$^3$-defect sites

Noémie Danné[1,2], Mijin Kim[3], Antoine G. Godin[1,2], Hyejin Kwon[3], Zhenghong Gao[1,2], Xiaojian Wu[3], Nicolai F. Hartmann[4], Stephen K. Doorn[4], Brahim Lounis[1,2], YuHuang Wang[3,5] and Laurent Cognet[1,2,*]

[1]Univ. Bordeaux, Laboratoire Photonique Numérique et Nanosciences, UMR 5298, F-33400 Talence, France

[2]Institut d'Optique & CNRS, LP2N UMR 5298, F-33400 Talence, France

[3]Department of Chemistry and Biochemistry, University of Maryland, College Park, MD 20742, United States

[4]Center for Integrated Nanotechnologies, Materials Physics and Applications Division,, Los Alamos National Laboratory, Los Alamos, NM 87545 USA

[5]Maryland NanoCenter, University of Maryland, College Park, MD 20742, United States

*Laurent.cognet@u-bordeaux.fr.

**ABSTRACT**

The intrinsic near-infrared photoluminescence observed in long single walled carbon nanotubes is systematically quenched in ultrashort single-walled carbon nanotubes (usCNTs, below 100 nm length) due to their short dimension as compared to the exciton diffusion length. It would however be key for number of applications to have such tiny nanostructure displaying photoluminescence emission to complement their unique physical, chemical and biological properties. Here we demonstrate that intense photoluminescence can be created in usCNTs (~40 nm length) upon incorporation of emissive sp$^3$-defect sites in order to trap excitons. Using super-resolution imaging at <25 nm resolution, we directly reveal the localization of excitons at the defect sites on individual usCNTs. They are found preferentially localized at nanotube ends which can be separated by less than 40 nm and behave as independent emitters. The demonstration and control of bright near-infrared photoluminescence in usCNTs through exciton trapping opens the possibility to engineering tiny carbon nanotubes for applications in various domains of research including quantum optics and bioimaging.

The introduction of chemical or structural defects in carbon nanostructures is a powerful route to shape and expand their optical properties. For instance, in nanodiamonds, nitrogen vacancies generate stable photoluminescence leading to the realization of quantum light sources, sensors, and of promising biological applications[1-4]. More recently, the incorporation of fluorescent quantum defects in single-walled carbon nanotubes (SWCNTs) created the opportunity to enhance SWCNT near-infrared photoluminescence (PL) properties[5-8] thus generating a growing interest for the realization of single photon sources or bioimaging[1-4,6,9,10]. Indeed, pristine semiconducting SWCNTs display low luminescence quantum yield, since their photophysical properties[5-8,11] are mainly imposed by non-radiative mechanisms, associated with structural and environmental defects[12-16]. In this framework, exciton diffusion, which spans typically one to several hundreds of nanometers along the nanotube[13,17], plays a key role since an exciton scans the integrity of the nanotube backbone before it may eventually emit a photon. As a consequence, ultrashort nanotubes (usCNTs)—semiconducting SWCNTs with lengths significantly shorter than the exciton diffusion range—display dominant non-radiative exciton decay mechanisms at nanotube ends which are efficient PL quenching sites and thus imperceptible photoluminescence[18,19]. Due to their exceptional small sizes and near-infrared optical properties, luminescent usCNTs would however be a key asset for a variety of applications including nano-electronics[20], biology[21,22] and nanomedicine[23].

In sp$^3$-defect functionalized SWCNTs, evidences for exciton localization at defects were obtained[5,24-26] and wide-field PL imaging showed emission from isolated defect sites[27]. However, the extent of localization could not be resolved due to the limited spatial resolution of standard microscopy (~ 450 nm for high numerical aperture (NA) objectives and emission wavelength of ~1100 nm) as compared to exciton diffusion lengths. Here we demonstrate that sp$^3$-defect functionalization can be used to trap excitons at emissive defect sites that are intentionally incorporated into usCNTs, and thus efficiently brightens usCNTs so that single nanotube detection of usCNTs becomes possible by fluorescence microscopy. Through the application of super-resolution imaging, we further reveal localization of emitting sites at <25 nm resolution on single usCNTs, well below the diffraction limit, usCNT lengths and free exciton diffusion lengths. Through this direct visualization, we could then demonstrate that defect sites are preferentially localized at usCNTs ends, and display independent emission properties.

We prepared length-sorted CoMoCAT usCNTs with median length of 43 nm following a previously published procedure[19]. The ultrashort lengths were determined by atomic force microscopy (AFM) measurement (N=88, **Figure 1a**) and single particle photothermal imaging (PhI). The photothermal signals are directly linked to nanoparticle absorption so that this distribution mirrors the AFM length distribution albeit with better statistics[19] (N= 472, **Figure 1b**). Ensemble absorption spectra confirmed the predominant presence of (6,5) nanotubes in the sample as expected from SG65i CoMoCAT nanotubes (Supporting Information **Figure S1**). On the other hand, an excitation-emission 2D PL map displayed imperceptible PL signals (**Figure 1c**) which is not surprising as it has been well established that usCNTs do not fluoresce[18,19,28]. Consistent with the negligible PL in the bulk samples, no luminescent usCNTs were detected at the single nanotube level even using long integration times (several seconds, **Figure 1d**).

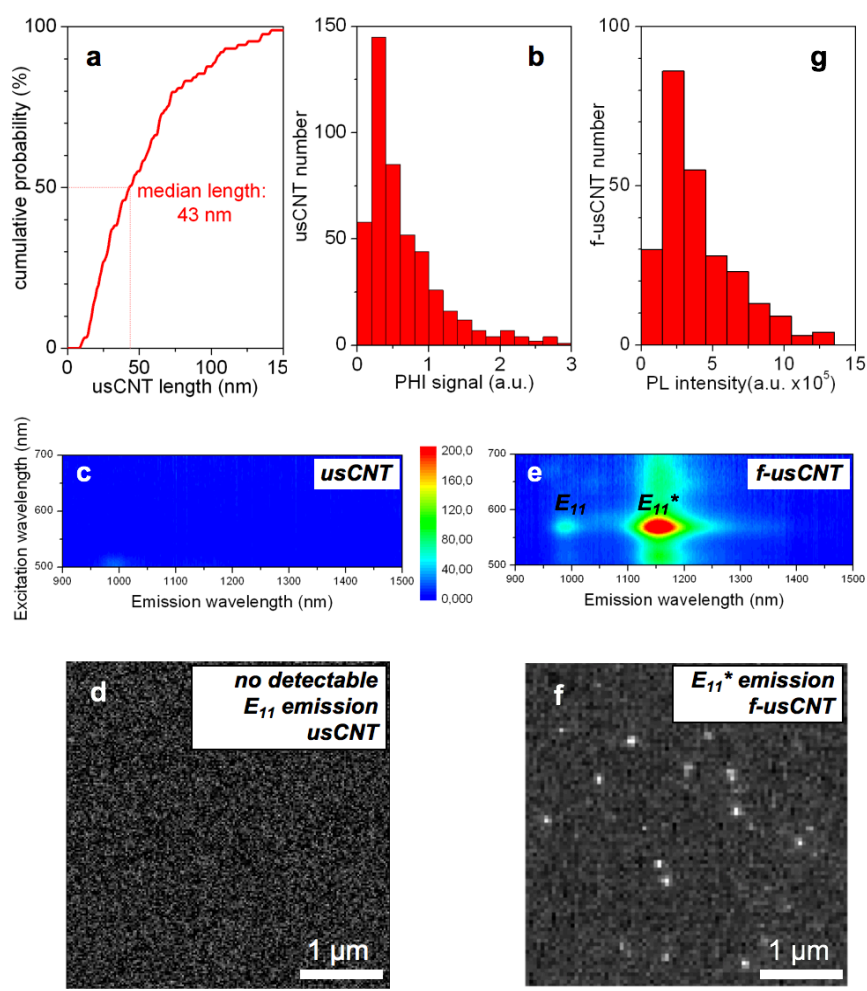

*Figure 1. Brightening of ultrashort SWCNTs through fluorescent quantum defects.* (a) Cumulative distribution of usCNT lengths measured by AFM (N=88, median= 43 nm, $1^{st}/3^{rd}$ quartile= 24/69 nm). (b), Distribution of usCNT absorption signals measured by PhI (N=472) which also reflects the nanotube lengths distribution but with better statistics. (c) 2D PL map of usCNTs shows no luminescence. (d) single molecule imaging of usCNTs is not observed. (e) 2D PL map of f-usCNT reveals bright luminescence. (f) single molecule imaging of f-usCNT is clearly observed. (g) Distribution of f-usCNT PL signals (N=265) mirrors nanotube length distributions in (a) and (b).

We next covalently attached perfluorinated-hexyl chain (-$C_6F_{13}$) to the usCNTs producing functionalized usCNTs (f-usCNTs). A 2D PL map of the f-usCNTs solution reveals a bright and stable PL peak, $E_{11}$* around 1160 nm (**Figure 1e**). In $C_6F_{13}$-functionalized, long (6,5)-SWCNTs, this red-shifted peak ($E_{11} - E_{11}$*, 183 meV) has been attributed to PL arising from the chemical functionalization[8]. Interestingly, weak yet photostable $E_{11}$ exciton emission also became observable from f-usCNTs, in contrast to the unfunctionalized usCNT samples. f-usCNTs that were spin-coated on a microscope glass coverslip pre-coated with polyvinylpyrrolidone were next observed at the single tube level. We used a dual detection fluorescence microscope in order to simultaneously detect the $E_{11}$ and/or $E_{11}$* PL bands of (6,5) f-usCNTs upon 568 nm excitation (Supporting Information **Figure S2**). Unambiguously, bright f-usCNT PL spots could be observed in the $E_{11}$* channel at the video frame rates (**Figure 1f**). All detected spots were diffraction limited as expected from single nanoscale emitters. In addition, we built a distribution of intensities measured from 265 luminescent spots and obtained a monomodal distribution (**Figure 1g**) in agreement with usCNT length and PhI signal distributions (**Figure 1a-b**). These observations reveal that the detected spots stem from individual f-usCNTs and thus that luminescent usCNTs are bright enough to be detected at the single tube level owing to $C_6F_{13}$-functionalization. Noteworthy, on occasion a few dim PL spots were detected in the $E_{11}$ channel and co-localized with $E_{11}$* PL spots from single f-usCNTs (Supporting Information **Figure S3**) confirming the observation in **Figure 1e**.

Because usCNT lengths (median length of 43 nm, **Figure 1a**) are significantly shorter than the exciton diffusion length (≥~ 100 nm), the efficient PL of f-usCNTs is an unambiguous signature of $E_{11}$* exciton localization (**Figure 2**). Indeed, the prevailing interpretation for the vanishing $E_{11}$ PL in usCNTs is that excitons systematically decay non-radiatively at nanotube ends acting as efficient quenching defects[15,18,29,30] (**Figure 2a**). $C_6F_{13}$ functionalization thus prevents diffusion-related end-quenching of $E_{11}$ excitons through the creation of localized $E_{11}$* excitons (**Figure 2b**). Interestingly, the observation of weak yet stable $E_{11}$ PL in f-usCNTs in contrast to usCNTs might indicate that thermal detrapping of $E_{11}$* excitons can occur from localized sites.

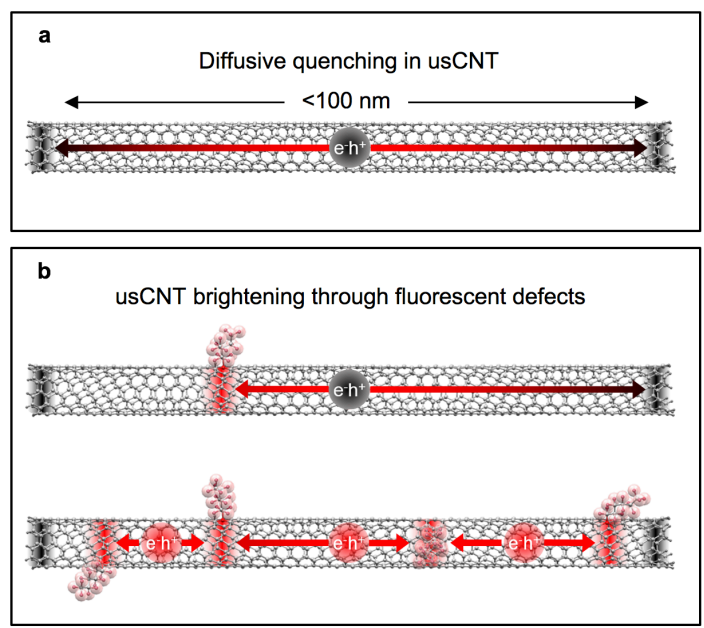

*Figure 2. Schematics of photo-excited $E_{11}$ exciton fate in usCNTs.* (a) exciton diffusion and quenching in unfunctionalized usCNTs. In an usCNT, exciton decay is not diffusion-limited but nanotube-length-dependent. End quenching imposes extremely low fluorescence quantum yields of usCNTs. (b) Schematics of exciton diffusion and trapping at defects in f-usCNTs. Mobile excitons in an f-usCNT can be trapped at fluorescent quantum defects and therefore efficiently luminesce (top: one luminescent defect, bottom: multiple defects at different positions).

In order to investigate the detrapping efficiency, we measured the temperature dependence of the ratio of $E_{11}$ PL intensity to that of $E_{11}^*$ PL measured on bulk spectra. This ratio follows an Arrhenius type relation (**Figure 3a**) with a fitted thermal detrapping energy of 109 meV, which is smaller than the optical gap (183 meV). This energy difference can be understood in terms of the vibrational reorganization energy associated with defect sites[24]. From the Arrhenius low, we estimated that ~2 % of the $E_{11}^*$ excitons can be transferred to the $E_{11}$ population through thermal detrapping at room temperature (**Figure 3b**). However, thermal detrapping cannot fully account for the total observed $E_{11}$ PL in the f-usCNTs spectra,. Another mechanism could be that the formation of trapped $E_{11}^*$ excitons from photo-created $E_{11}$ excitons requires crossing a small potential barrier[25] (**Figure 3c**). Finally, all these effects might be reinforced by an $E_{11}^*$ state filling effect previously suggested[31]. Indeed, we observed at the single nanotube level that in f-usCNTs, $E_{11}^*$ PL saturates at lower laser intensities than $E_{11}$ PL measured on longer SWCNTs (120 nm median length, Supporting Information **Figure S4**) prepared from the same batch as f-usCNTs (**Figure 3 d-e**). Saturation at lower intensities is also consistent with the observation of longer PL decays in sp$^3$ defect functionalized SWCNTs as compared to unfunctionalized SWCNTs[26]. Efficient $E_{11}^*$ state filling would then enhance the subpopulation of $E_{11}$ excitons in f-usCNTs and thus the observed $E_{11}$ emission (**Figure 3f**). We note that in all these three scenarios (**Figure 3d-f**), the presence of at least two defect sites is required to compete with diffusive quenching at the nanotube ends (**Figure 2b**).

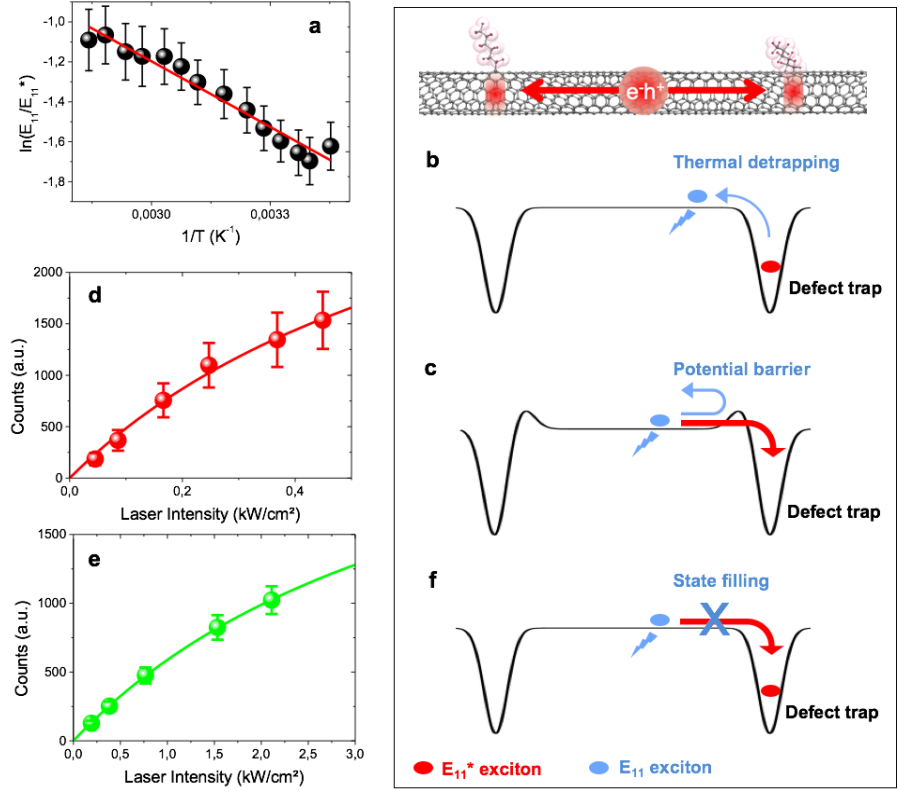

*Figure 3. Trapping and detrapping of $E_{11}^*$ excitons. (a) The van 't Hoff plot for f-usCNTs as derived from the integrated PL intensity ratio of E11 and E11\* at the corresponding temperature (black dots). The slope of the linear fitting (red line) provides the detrapping energy of 109 meV. (b) Schematic representation of thermal detrapping of a $E_{11}^*$ exciton to generate a mobile $E_{11}$ exciton. (c) Schematic representation of $E_{11}^*$ exciton trapping in the presence of a potential barrier leading to the presence of a small population of $E_{11}$ excitons. (d) Mean $E_{11}^*$ PL signal of individual f-usCNTs as a function of cw laser intensity (N = 8 usCNTs; circles: experimental data; solid line: fit using a saturation profile yielding a saturation intensity of 0.8 kW/cm². (e) Mean $E_{11}$ PL signal of individual long CNTs from the same preparation as usCNTs as a function of cw laser intensity (N = 12 long nanotubes; circles: experimental data; solid line: fit using a saturation profile yielding a saturation intensity of 4.3 kW/cm²). (f) Schematic representation of $E_{11}^*$ exciton state filling leading to the presence of a small population of $E_{11}$ excitons.*

Conversely, by preventing $E_{11}^*$ excitons from reaching and quenching at the nanotube ends, a single isolated site of $sp^3$-defect localization can in principle brighten usCNTs through emission at 1160 nm, while the incorporation of several sites, including at the usCNT ends, should further enhance f-usCNT $E_{11}$* photoluminescence (**Figure 2b**). To determine the relative localization of the defects along the usCNTs, the position of emitting sites separated by $\Delta d < 43$ nm must be imaged at a resolution much below the diffraction limit (1,22 $\lambda$/2NA ~ 500 nm, for $\lambda = 1160$ nm). This requires a fluorescence imaging modality having at least a ten-fold resolution improvement over standard diffraction-limited fluorescence microscopy. We also wish to investigate whether luminescent defects behave as independent emitters in f-usCNTs or whether their close proximity on the nanotube backbone would condition their PL properties.

To achieve the required resolution and considering that f-usCNT emission occurs in the near-infrared (i.e. at long wavelengths), we implemented a super-resolution microscopy strategy based on single emitter localization for its adequacy with f-usCNT photophysics — several super-resolution imaging techniques have been widely used in biology to achieve sub-wavelength resolution imaging in fluorescence microscopy[32]. Localization microscopy was previously demonstrated on pristine SWCNTs where the nanotubes were exposed to quenching moieties[33]. The presence of local extrinsic PL, defects were localized as missing PL signals at sub-wavelength resolution which also provided a direct visualization of the ~100 nm exciton excursion range in pristine SWCNTs. Yet, neither usCNTs which do not fluoresce nor defect tailored SWCNTs were super-resolved previously.

In the following, f-usCNTs were directly spincoated on glass coverslips and single f-usCNTs PL imaging was performed. In these conditions, the majority of nanotubes showed blinking behavior as similarly observed in long functionalized SWCNTs[27]. **Figure 4a-d** displays PL intensity traces acquired over several minutes revealing the frequent presence of multiple intensity levels for single f-usCNTs. This is quantified by analyzing the amplitude of 64 intensity steps on different f-usCNTs, presented in the distribution of **Figure 4e**. The observed distribution of intensity steps is multimodal, with a factor of two separating the two sub-populations that we observe, demonstrating that individual sites are imaged in the first subpopulation. Importantly, the presence of the second blinking population indicates that the PL behaviors from two defect sites can be uncorrelated on single (sub 50 nm) f-usCNTs. To unambiguously establish the independent localization of defect sites at a scale on f-usCNTs and rule out the possibility that nanotube bundles or outliers in nanotube length distribution were at the origin of the second intensity step population, defect localization was investigated by super-localization analysis on the f-usCNTs that displayed multiple intensity steps.

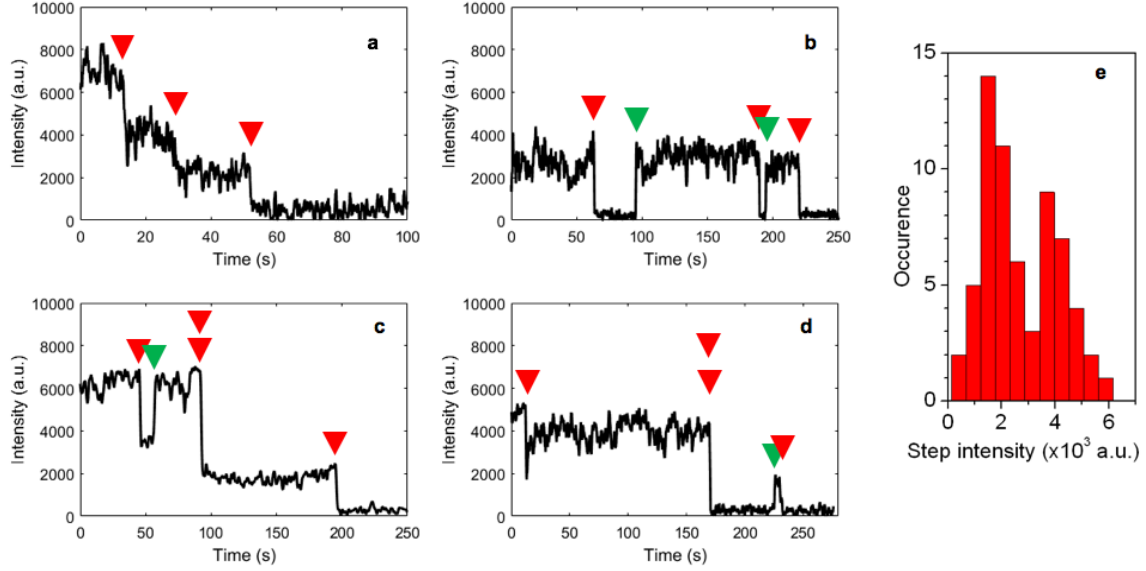

*Figure 4. Evidence of $E_{11}^*$ exciton localization*. (a-d) $E_{11}^*$ PL intensity traces of 4 different single blinking f-usCNTs as a function of time. Red (resp. green) triangles represent negative (resp. positive) unitary steps. (e) Histogram of $E_{11}^*$ PL step intensities observed in single blinking f-usCNTs (N=64). The histogram is multimodal which reveals the presence of several individual emitting sites in f-usCNTs. The presence of independent intensity levels in f-usCNTs having lengths shorter than the diffusion length of $E_{11}$ excitons in long CNTs is a signature of $E_{11}^*$ exciton.

For each intensity step event, differential images were computed, consisting of the differences between successive images before and after the step, in order to isolate the appearance or disappearance of the PL stemming from an individual luminescent site (**Figure 5a** and Supporting Information **Figure S5**). When steps were negative, the absolute value of the pixel intensities in the differential image was considered to obtain images with positive pixel values. Images preceding and following the intensity step over time are averaged in order to improve signal-to-noise ratio in the differential images and subsequent localization precision (see Supporting Information methods and **Figure S5**). Individual spots in differential images were then fitted by a Gaussian approximating the point-spread function as commonly performed in localization microscopy[34]. Indeed, these fits retrieve the locations of individual emitters with sub-wavelength accuracy[35]. Doing so, maps of defect PL localization were created for several f-usCNTs showing more than one intensity level (i.e. at least two independent emitting sites). When background noise can be neglected relative to the number of detected photons $N$, the localization precision can be approximated by $\sim s/N^{1/2}$, where $s$ is the standard deviation of the point spread function. In practice, camera pixelization and noise should be taken into account[36] (see Supporting Information methods) leading to localization precisions that could be estimated for each individual defect site. Taking into account localization precision, on each single tube, defect detections that were below the localization precision were considered as one single defect (see Supporting Information methods and **Figure S6**). The mean localization precision for individual defect sites was ultimately determined to be 22 nm in our experiments (see Supporting Information methods). Subsequent defect localizations are displayed for several f-usCNTs, revealing the presence of at least two unmistakably resolved defects on **Figure 5a-e** with localization precision displayed as dotted circles.

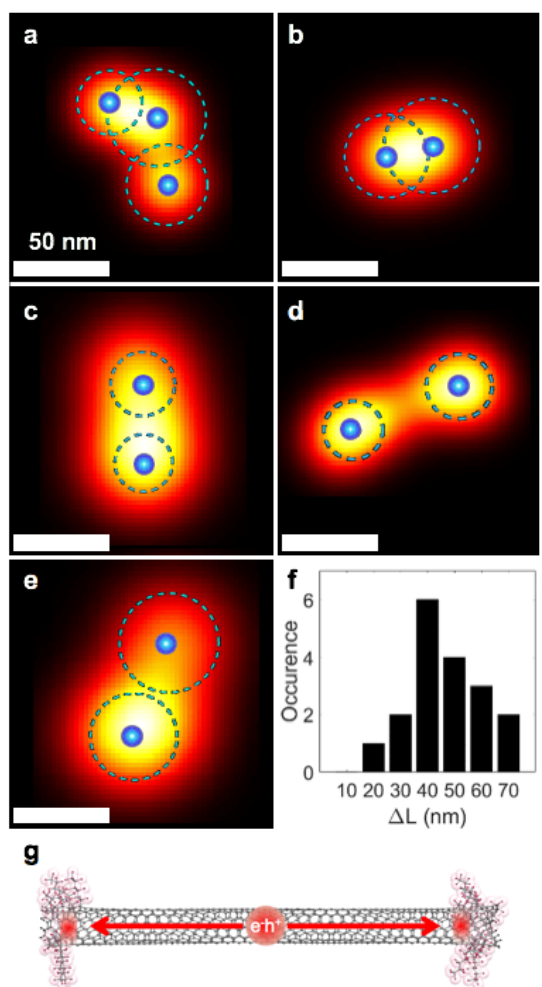

*Figure 5. Super-resolved images of $E_{11}*$ exciton localization in f-usCNTs. (a-e) Image representation of single blinking $E_{11}*$ exciton localizations in f-usCNTs by colored 2D Gaussian curve rendering. For this, each localization is displayed as a two-dimensional Gaussian of unit amplitude and width equal to the localization precision as commonly used in localization microscopy[34]. The mean $E_{11}*$ exciton localization of each single site is represented by blue spots and the corresponding localizaton precision by dotted circles (see supporting information methods and Figure S5). (f) Histogram of super-resolved distances ΔL between $E_{11}*$ exciton localizations (18 f-usCNTs analyzed). (g) Schematic of defect sites preferentially localized at nanotube ends.*

These images constitute a direct visualization of exciton localization at defect sites to locations below the measured precision, which is significantly smaller than the exciton diffusion range in pristine nanotubes. Interestingly, the distribution of distances between independently localized defects supports this finding (**Figure 5f**), but also indicates that typical distances between defects match the f-usCNT lengths (**Figure 1a-b**) and PL signals (**Figure** 1f). This observation indicates that functionalization preferentially occurred at nanotube ends (**Figure 5g**). We suspect that the functionalization occurs propagatively from the end defects,[37,38] creating fluorescent quantum defects[7,8] that block excitons from being trapped by the otherwise quenching ends.

In this work, we achieved dramatic PL brightening of ultrashort carbon nanotubes upon defect functionalization, which otherwise do not luminesce. Brightening is so efficient that ultrashort carbon nanotubes can be detected at the single nanotube level. This brightening of sub-100 nm nanotubes is a direct consequence of exciton localization at intentionally introduced defects in the functionalized tubes. To directly support this mechanism, we developed a super-resolution imaging methodology to resolve the different emission sites with <25 nm resolution on single tubes. Super-resolution imaging unambiguously reveals that emitting excitons are localized at nanotube ends. Brightening of ultrashort nanotubes not only constitutes a direct evidence for exciton localization in defect functionalized tubes, it also opens a promising route for several applications based on carbon nanotubes. For instance, luminescent ultra-short nanotubes with a controlled number of single emitters might be highly valuable for the development of high quality single photon emitters for quantum information[10]. We also foresee that luminescent ultra-short nanotubes will constitute a milestone for biological imaging where ultra-small, bright photostable emitters in the near-infrared biological window are vividly required[39].

**ACKNOWLEDGEMENTS**

This work was supported by CNRS, the French Ministry of Higher Education, Research and Innovation, the Agence Nationale de la Recherche (ANR-14-OHRI-0001-01), IdEx Bordeaux (ANR-10-IDEX-03-02), and Conseil Regional Nouvelle-Aquitaine (2015-1R60301-00005204). YHW gratefully acknowledges the National Science Foundation of U.S. for financial support through grant no. 1507974 for the development of the defect chemistry. The authors thank having had access to the Center for Integrated Nanotechnologies, a U.S. Department of Energy Office of Science user facility.

# SUPPORTING INFORMATION

**METHODS**

***Ultrashort SWCNT preparation.*** usCNTs were prepared following the protocol described in ref[1]. In brief, 30 mg of raw CoMoCAT nanotubes were dispersed in 10 mL of milli-Q water containing 0.3 wt% sodium deoxycholate (DOC) by pulsed tip sonication in an ice bath (output at 45 W, pulse 0.5 s and pause 0.2 s) for 5 h. After sonication, gentle centrifugation (30 min, 3500 rpm, Eppendorf Centrifuge 5804 R) was performed to remove the majority of bundles and insoluble materials. The supernatant (9 mL) was diluted to 20 mL with 0.3 wt% DOC solution and an ultracentrifugation was performed to remove any remaining bundles and catalyst particles (45000 rpm, i.e.184000 g, 2 h, Thermo Scientific Sorvall WX+ Ultracentrifuge). This process was repeated twice, and the supernatant was collected for nanotube length separation by density gradient ultracentrifugation (DGU)[2]. A density gradient solution was prepared via subsequently stacking four layers of 5 mL of 60 wt%, 10 wt%, 7.5 wt%, and 5 wt% iodixanol solution (OptiPrep, Sigma-Aldrich) in 0.3 wt% DOC solution, from bottom to top in a polycarbonate centrifugal tube (Beckman Coulter) with a volume of 26.3 mL. Iodixanol (60 wt%) was diluted in 0.3 wt% DOC solution to reach required concentrations. A layer of 5 mL freshly prepared sonication-cut SWCNT suspension was placed on top of the density-gradient iodixanol solution and centrifuged at 184,000g for 3h at 4 °C. After ultracentrifugation, SWCNTs with different lengths were distributed along the density gradient with the shortest usCNTs near the top. Fractioned samples were collected from top to bottom. Their lengths were characterized by AFM and photothermal imaging (**Figure** 1). For long-term storage, the concentration of DOC in all received samples was raised to 1 wt% to maintain individual dispersion of nanotubes.

***Covalent functionalization of usCNTs.*** The usCNTs were covalently functionalized by perfluorinated hexyl groups[3]. ($-C_6F_{13}$)[3]. To an aqueous solution of usCNTs in 1% w/v sodium dodecyl sulfate-$D_2O$, 7.6 mM of $NaHCO_3$ (EMD chemicals, HPLC grade), 0.16 %v/v $CH_3CN$ (Acros organics, HPLC grade, 99.9 %) and perfluoro hexyl iodide (Sigma Aldrich, 99%) were added. 3.6 mM of $Na_2S_2O_4$ (Sigma Aldrich, 85 %) was then added to the mixture and stirred at room temperature with protection from ambient light. After 24 hours of reaction, the SWCNT solutions were characterized by absorption and fluorescence spectroscopy. The absorption spectra were taken on a Lambda 1050 UV-vis-NIR spectrophotometer (Perkin Elmer) equipped with both a photomultiplier tube and an extended InGaAs detector. The PL of f-usCNT solutions was collected using a NanoLog spectrofluorometer (HORIBA Jobin Yvon). The samples were excited with a 450 W Xenon source dispersed by a double-grating monochromator. The slit width of the excitation and emission beams were 10 nm. The PL spectra were collected using a liquid-$N_2$ cooled linear InGaAs array detector.

*Near-infrared single nanotube PL imaging.* Single SWCNT PL imaging was performed with an inverted microscope equipped with a 1.40 NA 60x objective and two detection arms separated by a 50/50 beam splitter (see Supplementary **Figure** S2). SWCNTs were excited in a widefield geometry with a 568 nm cw solid-state laser (Coherent Saphirre) at 0.5 kW/cm² otherwise stated (**Figure** 3). The excitation beam was circularly polarized light. A dichroic mirror (FF875-Di01, Semrock) in combination with a long-pass emission filter (RazorEdge 1064, Semrock) for (6,5) functionalized SWCNTs was used to illuminate and select $E_{11}$* PL. $E_{11}$ PL of (6,5) SWCNTs was detected with an emCCD camera (Princeton Instruments ProEM), while $E_{11}$* PL of (6,5) functionalized SWCNTs was detected by an InGaAs camera (XEVA 1.7 320 TE3). For imaging, usCNTs were suspended in 1% w/v sodium dodecyl sulfate (SDS) and spin-coated on microscope slides at 3000 rpm for 3 minutes.

*Photoluminescence intensity traces of blinking f-usCNTs.* Movies of nanotubes blinking were recorded for up to 2000 images (integration times of 100 to 200 ms per frame). Intensity traces correspond to the integrated intensity of a given nanotube spot in the movie frames. For display, intensity traces were smoothed by the Savitsky-Golay filter to increase signal-to-noise ratios. Intensity steps displayed in **Figure** 4e were calculated by 2D Gaussian fitting of the spots appearing in image differences corresponding to the subtraction of images immediately before and after the step. Intensity steps correspond to the integral of the Gaussian fits using the software ImageJ.

*Super-resolution imaging of $E_{11}$* excitons in f-usCNTs.* In order to localize $E_{11}$* excitons in f-usCNTs, f-usCNTs displaying PL blinking were analyzed using the following procedure (see Supplementary **Figure** S5): (i) N images immediately before and after an intensity step were averaged to enhance signal-to-noise ratio (N varied between 5 and 30 depending on intensity steps). (ii) the resulting averaged images were subtracted to create a differential image isolating a blinking event from one emitting defect. The absolute PL values are taken into account in case of a positive blinking event to equally analyze positive and negative steps. (iii) Each spot in a differential image was then fitted by a 2D Gaussian profile with the plug-in GDSC SMLM (ImageJ), as used in single molecule localization microscopy,[4-6] to retrieve the emitter localization with sub-diffraction precision. (iv) for visualization (**Figure** 5 and Supplementary **Figure** S6b-c), super-resolved images are produced from the individual defect localizations displayed as color-coded two-dimensional Gaussians of unit amplitude and width equal to the precision of localization. Localization precision of each event was calculated taking into account signal-to-noise ratio of the blinking event fitted by the 2D Gaussian at step (iii) in the initial differential image, as well as the pixel size in the image as described in ref. [7]. In Supplementary **Figure** S6c, each individual defect detection is also displayed as a red spot. (v) In order to identify individual emitting defects that might be detected several times through multiple blinking events, adjacent detections that were situated within the range of neighboring localization precisions (white dotted circle in Supplementary **Figure** S6d) were merged into one single defect at the barycenter position (blue dot in Supplementary **Figure** S6e) and displayed with the corresponding localization precision in final images (blue circles in **Figure** 5a-e and Supplementary **Figure** S6f). The average localization precision was equal to 22 nm.

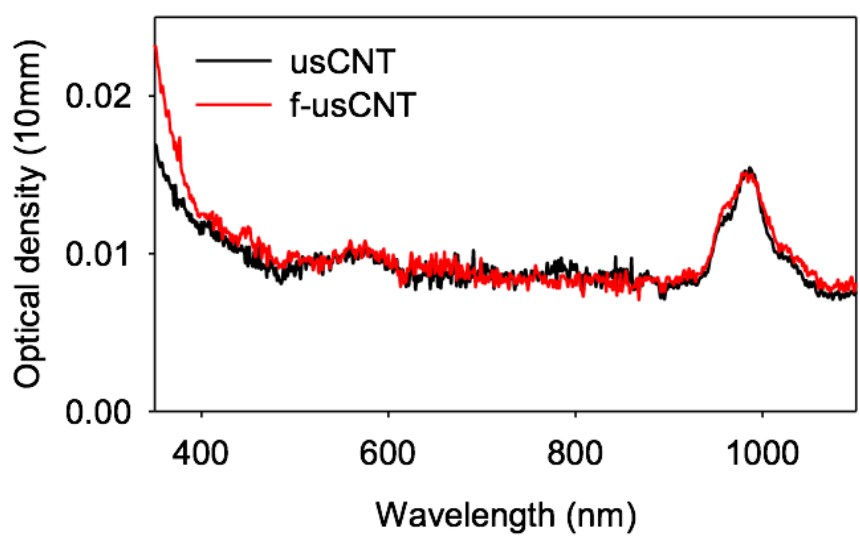

*Figure S1 | The absorption spectra of 43 nm long SG65i CoMoCAT usCNTs* before (black) and after functionalization by perfluorinated hexyl defects (red).

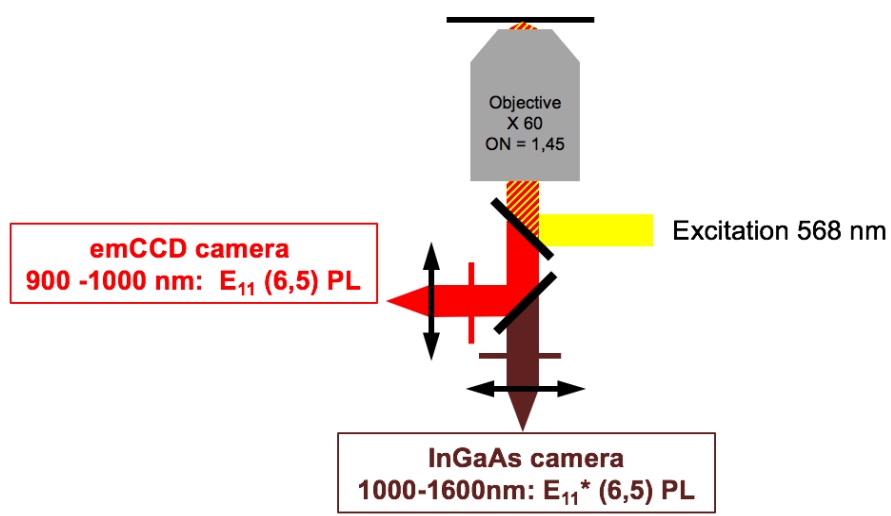

*Figure S2 | Single nanotube luminescence microscope.* Dual-detection single nanotube microscope for the detection of $E_{11}$ and $E_{11}^*$ PL emission of (6,5) f-usCNTs. The spectral bands of each detection arm are indicated.

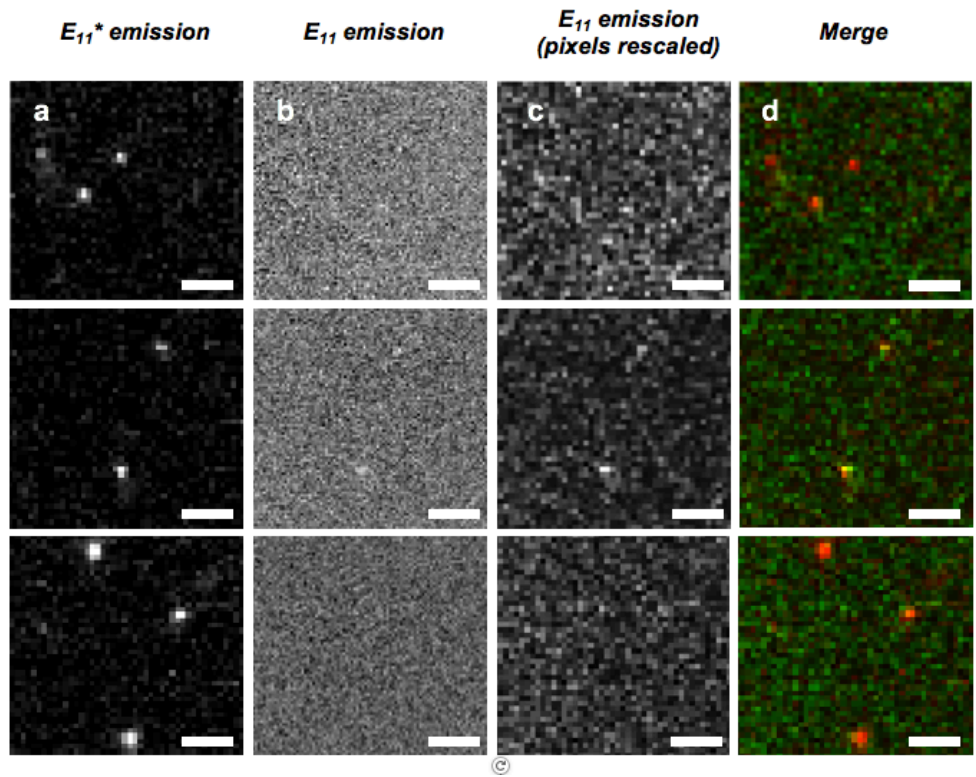

***Figure S3 | Dual-detection of single (6,5) f-usCNTs*** in $E_{11}^*$ (***a***) and $E_{11}$ (***b***) *PL detection channels. On occasion (e.g. second row), weak $E_{11}$ PL is detected and is colocalized with $E_{11}^*$ PL as shown in composite images in (**d**). Red: $E_{11}^*$ and green: $E_{11}$ channels. For creating composite images in (**d**), images from the $E_{11}$ channel have been rescaled in (**c**) to match pixel sizes of images in (**a**). Integration times: 200 ms. Scale bar = 5µm.*

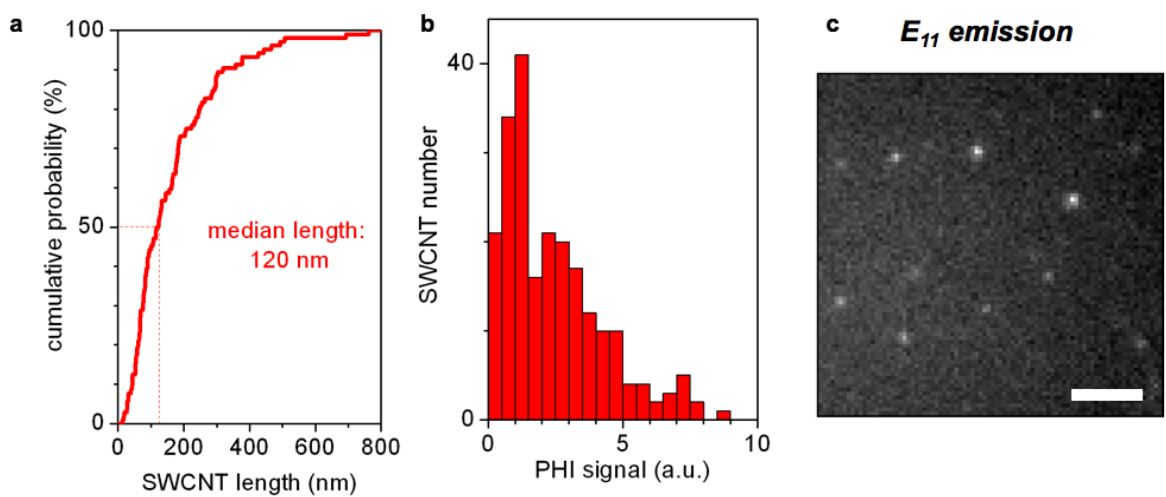

*Figure S4 | Characterization of the SWCNTs used in Fig. 3e. **a**, Cumulative distribution of SWCNT lengths measured by AFM (N=104). **b**, Distribution of SWCNT absorption signals measured by PhI (N=244) which also reflects the SWCNT length distribution. **c**, $E_{11}$ PL image of single SWCNTs. Integration time: 200ms. Scale bar: 5µm. See Fig. 1a,b&d for comparison to usCNTs.*

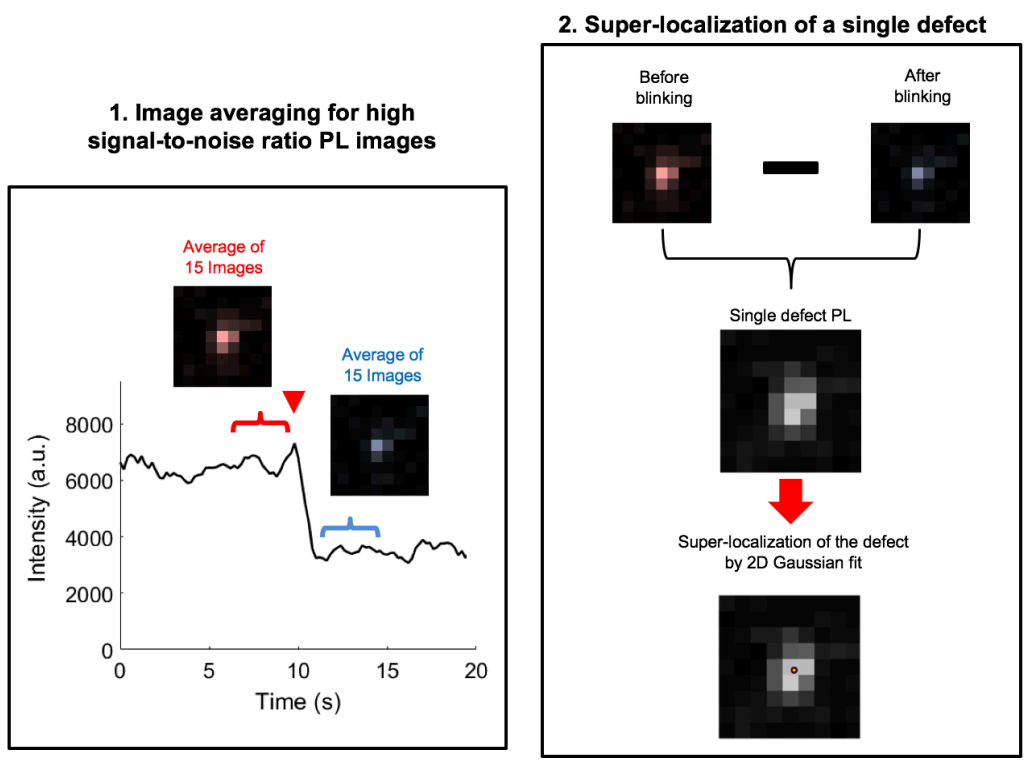

*Figure S5 | Super-localization of a single defect (see methods).*

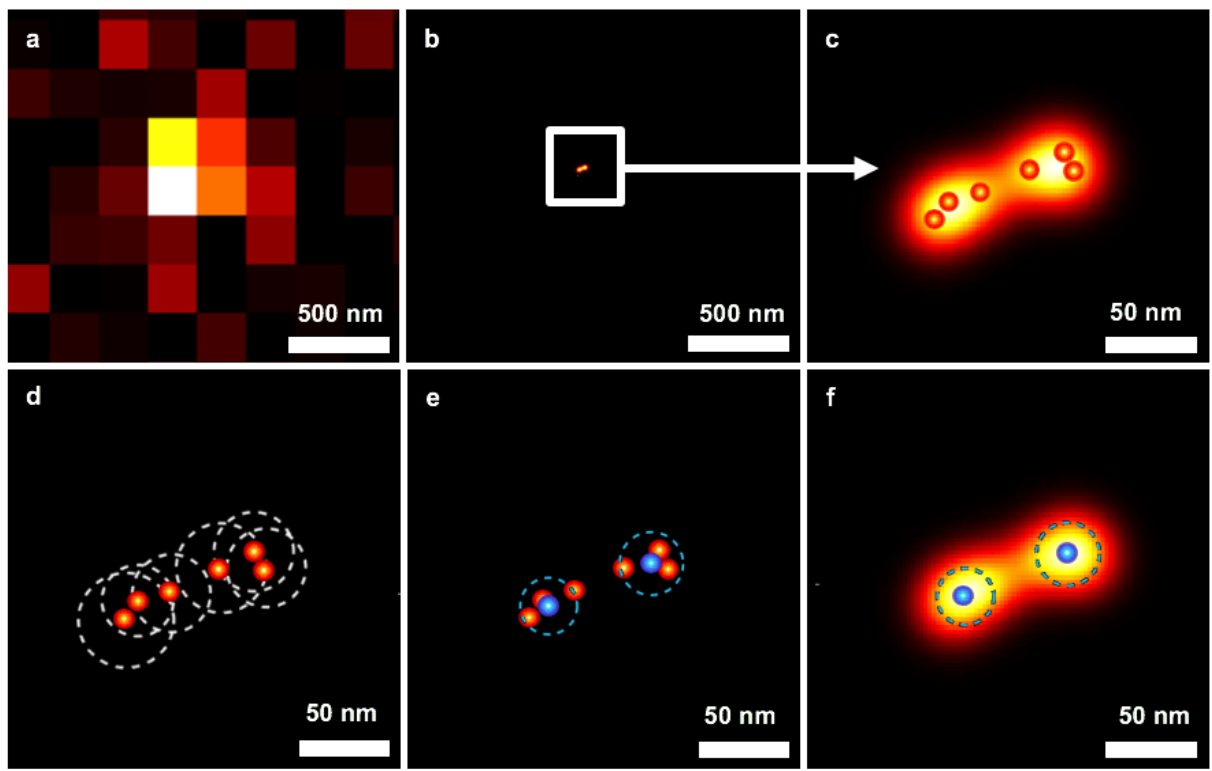

*Figure S6 | Super-resolved imaging of $E_{11}^*$ exciton localization in f-usCNTs. a,* Differential image revealing a blinking event from $E_{11}^*$ PL emission of a single (6,5) f-usCNT. *b,* Corresponding super-resolved image produced from 6 individual defect localizations displayed as color-coded two-dimensional Gaussian distributions of unit amplitude and widths equal to the precision of each localization. *c,* Same as b. but with zoomed scale. The localization of each of the 6 detections is indicated by a red spot. *d,* The 6 localizations are represented with the precision of their localization. *e,* The blue dots represent the two emitting defect sites at the origin of the blinking events detected in a-d. The localization precisions of the emitting defects are indicated by blue circles. *f,* Super-resolved image of the two defect sites with their localizations as revealed by the procedure presented in a-e.